# Reliable Probabilistic Gossip over Large-Scale Random Topologies

Ruijing Hu[1] and Leander Jehl[2]

[1]College of Informatics, Huazhong Agricultural University, China
[2]Department of Electrical Engineering and Computer Science, University of Stavanger, Norway

**Abstract**

This paper studies reliability of probabilistic neighbor-aware gossip algorithms over three well-known large-scale random topologies, namely Bernoulli (or Erdős-Rényi) graph, the random geometric graph, and the scale-free graph. We propose a new and simple algorithm which ensures higher reliability at lower message complexity than the three families of gossip algorithms over every topology in our study.

We also present a uniform approach to model the reliability of probabilistic gossip algorithms in the different random graphs, whose properties, in fact, are quite different. In our model a forwarding probability is derived with consideration of parameters in gossip algorithms and graph properties. Our simulations show that our model gives a reasonable prediction of the trade-off between reliability and message complexity for all probabilistic neighbor-aware gossip algorithms in various random networks. Therefore, it allows to fine-tune the input parameters in the gossip protocols to achieve a desirable reliability with tolerable message complexity.

## I. INTRODUCTION

In the past decades, information dissemination has become an essential issue in large-scale networks, where a global view is often not available for distributed applications [14]. In order to reach all sites, even without a direct connection, a source has to rely on other sites to successively retransmit the message across the network. For example, in an RSS delivery system that is considered reliable, an updated stream from any publisher should notify every site when the dissemination ends.

A straightforward solution to this problem is using a *pure flooding* protocol [25], where upon the first reception of a message, a site forwards it to all its neighbors. Evidently, sites with more than two neighbors will receive many redundant copies of the message. It might even give rise to a *broadcast storm* [34], which can easily reveal bottlenecks, entailing for instance network congestion and message loss. The reliability is thus degraded.

Aiming to mitigate these drawbacks, many optimized *gossip protocols* have been proposed, which require the sites of the system to relay a message only to some of their neighbors. The retransmission scheme is either carried out in a deterministic or probabilistic way. The former can provide high reliability, but its implementation is commonly very hard (i.e. in some schemes it is proven to be NP-hard to reach the optimal performance) and requires more information [30]. Probabilistic schemes stand out through their simplicity and scalability [17], [42], while their reliability highly depends on the choice of scheme and pre-configured input parameters. In particular, probabilistic neighbor-aware gossip algorithms specify the parameters and forwarding decision for every site with regard to the local view about their neighbors, which can leverage the performance [24].

In this article, we study three basic probabilistic neighbor-aware gossip algorithms from the literature: (1) Fixed Fanout Gossip (*GossipFF*) [26], (2) Probabilistic Edge Gossip (*GossipPE*) [40], and (3) Probabilistic Inverse Self-degree Broadcast Gossip (*GossipPISB*) [9]. Furthermore, we propose a reliable algorithm: Probabilistic Inverse Neighbor-degree Edge Gossip (*GossipPINE*), which takes advantage of local information about sites in a one-hop neighborhood. In terms of reliability, it performs better than the basic gossip algorithms over three widely-studied random graphs: Bernoulli (or Erdős-Rényi) graph ($\mathcal{B}(N, p_N)$) [11], the Random geometric graph ($\mathcal{G}(N, \rho)$) [4], [36], and a scale-free graph generated by Barabási-Albert model ($\mathcal{S}(N, m)$) [3], which respectively model peer-to-peer system overlays [26], wireless sensor networks [40], and social networks [19].

We model the reliability as the probability of isolated site in a dissemination graph. We thus describe both reliability and message complexity based on the forwarding probability between neighbors. This forwarding probability can be adjusted to parameters in the gossip algorithms and graph properties, like edge dependency and the preferential attachment in the Barabási-Albert model. Our model is applied to all algorithms on every topology, and gives reasonable predictions of trade-off between reliability and message complexity. To the best of our knowledge, it is the first time that such a general model for the reliability of gossip algorithms is presented and evaluated. Interestingly, we find that our algorithm *GossipPINE* is very simple to model and not affected by more complex features of the random topology, such as edge dependency and preferential attachment. Moreover, the message complexity needed to achieve high reliability of *GossipPINE* only varies moderately over the different topologies. Contrarily, the reliability of the other algorithms shows a significant variation from one topology to another. We therefore argue that *GossipPINE* is easy to deploy even in the random networks, whose global characteristics are unknown beforehand.

Our extensive simulations on top of OMNET++ [1] evaluate reliability of the studied algorithms as well as latency, which is another important metric for information dissemination. In comparison, our models mostly match the experiments, while the superior efficiency of our algorithm is confirmed.

## II. SYSTEM TOPOLOGIES

In the following, let $|l|$ denote the size of the set $l$, and $P_{connect}(s_i \sim s_j)$ the probability that a site $s_i$ is connected to a site $s_j$.

A network underlying a large-scale dissemination system can be viewed as a bidirectional or undirected graph. It is comprised of $N$ sites $\{s_1, s_2, \cdots, s_N\}$. The set of sites $s_j$, connected to $s_i$ (i.e., $s_i \sim s_j$), is called $s_i$'s neighbors, and denoted $\Lambda_i$. $V_i = |\Lambda_i|$ denotes the degree of $s_i$; $P(k)$ represents the degree distribution of sites with $k$ neighbors (i.e., the fraction of sites with degree $k$) in the graph and $\bar{V}$ is the mean degree $\left(\bar{V} = \sum_{k=0}^{N-1} P(k) \cdot k\right)$. Moreover, another important property of the graph is defined as follows.

**Edge Dependency (or Clustering Coefficient), denoted** $C$ of a given random graph, for distinct sites $s_i, s_j, s_k$, is the conditional probability that, given the existence of edges $s_i \sim s_k$ and $s_j \sim s_k$, an edge $s_i \sim s_j$ also exists (i.e., $P_{connect}(s_i \sim s_j | s_i \sim s_k, s_j \sim s_k)$).

Our study consists of the following random topologies: Bernoulli (or Erdős-Rényi) graph $\mathcal{B}(N, p_N)$ [11], the random geometric graph $\mathcal{G}(N, \rho)$ [36], and a scale-free graph $\mathcal{S}(N, m)$ [3]. These graphs model peer-to-peer system overlays in [26], wireless sensor networks in [21], and social networks such as Facebook and Twitter in [8] respectively.

**Bernoulli (or Erdős-Rényi) graph:** In $\mathcal{B}(N, p_N)$ every pair of sites is connected with probability $p_N$, independent of other pairs. Based on [10], we suppose that $p_N > \frac{(1+\varepsilon) \cdot \ln(N)}{N}$, with a positive constant $\varepsilon$, aiming at having a *giant component* with $N$ sites and Poisson-law degree distribution $P(k) = \exp(-\bar{V}) \frac{\bar{V}^k}{k!}$, where $\bar{V} = p_N \cdot N$. Since the existence of an edge over $\mathcal{B}(N, p_N)$ does not depend on other edges, $C = P_{connect}(s_i \sim s_j) = p_N$.

**The random geometric graph:** $\mathcal{G}(N, \rho)$ is a graph whose sites are positioned uniformly at random in a bounded region. In this article, such a region is a rectangular plane with length $a$ and width $b$. Furthermore, two sites are connected, whenever the distance between them is at most $\rho$. Based on [38], we can fine-tune $\rho > \sqrt{\frac{(1+\varepsilon) \cdot \ln(N) \cdot a \cdot b}{N \cdot \pi}}$ with a positive constant $\varepsilon$ in order to ensure that the graph is connected. When ignoring the impact of the border effect on the degree distribution in $\mathcal{G}(N, \rho)$, it follows a Poisson-distribution [27]: $P_s(k) = \exp(-\bar{V}_s) \frac{\bar{V}_s^k}{k!}$ with $\bar{V}_s = \frac{N \cdot \pi \cdot \rho^2}{a \cdot b}$.

Nevertheless, we have observed in our simulations that the sites with low degrees near the boundary have an important impact on reliability, which cannot be ignored. As shown in [22], the degree distribution is $P(k) = P_s(k) \left( \frac{(a-2\rho)(b-2\rho)}{ab} + \frac{2a\rho + (b-2\rho)\rho}{ab} \psi(k) \right)$ where $\psi(k) = \int_0^1 \exp\left(-\bar{V}_s(F(x)-1)\right) \cdot F(x)^k \mathrm{d}x$ with $F(x) := \frac{1}{\pi}\left(x\sqrt{1-x^2} - \arccos(x)\right) + 1$. Trivially, the mean degree can be calculated as $\bar{V} = \sum_{k=1}^{N-1} P(k) \cdot k$.

In [5], the edge dependency is calculated as $C = 0.5865$.

**Scale-free graph:** $\mathcal{S}(N, m)$ is a random graph generated by the Barabási-Albert model [3]. Starting from a small *clique* of $m_0$ sites, at every step a new site is added, and connected to $m$ ($\leqslant m_0 \ll N$) sites already present in the graph. The probability $p$ that a new site will be connected to an existing site is proportional to the degree of this site. This is called *preferential attachment*. This process ensures that the graph is connected with a power-law degree distribution approximately equal to $P(k) = \frac{2m(m+1)}{k(k+1)(k+2)}$, where $k \geqslant m$ and $\bar{V} = 2m$ [39]. In this network, there are **hub** and **periphery** sites which have degree greater than $2m$ and between $m$ and $2m$ respectively.

In [18], the edge dependency is derived as $C = \frac{m-1}{8} \frac{(\log N)^2}{N}$, if $N$ and $m$ are large. Notice that the edge dependency in $\mathcal{S}(N, m)$ is very low, almost in the same order as in $\mathcal{B}(N, p_N)$.

## III. Gossip Algorithms

Information dissemination in large-scale network is commonly studied on the basis of Algorithm 1. Initially, the source sends a message to *all* of its neighbors (Lines 2 and 3). A site delivers and retransmits a received message provided the site has not previously received it; otherwise the message is discarded. Sites that have received the message at least once are called **infected sites**.

Probabilistic neighbor-aware gossip algorithms in the literature can be classified into three families, based on their implementation of the **Gossip()** procedure: (1) Fixed Fanout Gossip (*GossipFF*) [26], (2) Probabilistic Edge Gossip (*GossipPE*) [40], and (3) Probabilistic Inverse

**Algorithm 1**: Generic Gossip Algorithm

```
1  Broadcast (⟨msg⟩)
2  |  foreach s_j ∈ Λ_i do
3  |  |  Send(⟨msg⟩, s_j)

4  Receive (⟨msg⟩)
5  |  if msg ∉ msgHistory then
6  |  |  Deliver(⟨msg⟩) ;
7  |  |  msgHistory ← msgHistory ∪ {⟨msg⟩} ;
8  |  |  Gossip(⟨msg⟩,parameters) ;
```

**Algorithm 2**: Fixed Fanout Gossip (at $s_i$)

```
9  /* fanout: number of selected
      neighbors                         */
10 GossipFF (⟨msg⟩,fanout)
11 |  if fanout ⩾ V_i then
12 |  |  toSend ← Λ_i
13 |  else
14 |  |  toSend ← ∅
15 |  |  for f = 1 to fanout do
16 |  |  |  random select s_j ∈ Λ_i/toSend
17 |  |  |  toSend ← toSend ⋃ s_j
18 |  foreach s_j ∈ toSend do
19 |  |  Send(⟨msg⟩, s_j)
```

**Algorithm 3**: Probabilistic Edge Gossip (at $s_i$)

```
20 /* p_e: probability to use an
      edge                              */
21 GossipPE (⟨msg⟩,p_e)
22 |  foreach s_j ∈ Λ_i do
23 |  |  if Random() ⩽ p_e then
24 |  |  |  Send(⟨msg⟩, s_j)
```

**Algorithm 4**: Probabilistic Inverse Self-degree Broadcast Gossip (at $s_i$)

```
25 /* c_v: constant control for
      broadcast                         */
26 GossipPISB (⟨msg⟩,c_v)
27 |  if Random() ⩽ min{c_v/V_i, 1} then
28 |  |  foreach s_j ∈ Λ_i do
29 |  |  |  Send(⟨msg⟩, s_j)
```

**Algorithm 5**: Probabilistic Inverse Neighbor-degree Edge Gossip (at $s_i$)

```
30 /* c_e: constant control to use
      an edge                           */
31 GossipPINE (⟨msg⟩,c_e)
32 |  foreach s_j ∈ Λ_i do
33 |  |  if Random() ⩽ min{c_e/V_j, 1} then
34 |  |  |  Send(⟨msg⟩, s_j)
```

Self-degree Broadcast Gossip (*GossipPISB*) [9]. Besides the received message, the Gossip() procedures takes one parameter which is identical for all sites.

In *GossipFF* (Algorithm 2), a site $s_i$ sends $msg$ to a fixed number of sites, denoted $fanout$, in $\Lambda_i$, which are randomly selected (Lines 15-17). If $fanout \geqslant V_i$, $s_i$ transmits $msg$ to all its neighbors (Lines 11 and 12). For $fanout \geqslant \max\{V_1, V_2, \cdots V_N\}$, Algorithm 2 becomes a pure flooding algorithm.

In *GossipPE* (Algorithm 3), a site $s_i$ chooses to send $msg$ over an edge independently from the other edges with a fixed probability $p_e$ (see Line 23). Note that when $p_e = 1$, we obtain the flooding algorithm. From Algorithm 3, Random() generates a random number in the interval $[0, 1]$.

In *GossipPISB* (Algorithm 4), every site except the source diffuses $msg$ to all its neighbors independently with probability $\min\{\frac{c_v}{V_i}, 1\}$ (see Line 27). In particular, when $c_v \geqslant \max\{V_1, V_2, \cdots, V_N\}$ this protocol is the flooding algorithm.

*Our Algorithm - GossipPINE:* The idea behind our algorithm is to forward the message with higher probability towards sites with low degree. This ensures that sites with very low degree are infected, if one of their neighbors is infected. On the other hand, fewer redundant message copies are received at the sites with high degree. This will be analyzed in Sections V and VI.

Algorithm 5 shows our Probabilistic Inverse Neighbor-degree Edge Gossip, denoted *GossipPINE*. We assume that every site knows the degree of its neighbors. Like *GossipPISB*, site $s_i$ randomly chooses its edges over which $msg$ should be transmitted. However, the probability to send on one edge depends on the degree of the connected neighbor (see Line 33).

## IV. PERFORMANCE METRICS

In the context of information dissemination, the following metrics are commonly used for performance evaluation [13], [26], [29], [31] :

**Message Complexity, denoted M** measures the mean number of messages received (or sent, since no message loss is taken into account) by each site. The Message Complexity is given by $M = \frac{\Omega}{N-1}$ where $\Omega$ is the total number of messages exchanged during the dissemination.

**Reliability, denoted R** is defined as the percentage of messages generated by a source, that are delivered by all sites. A reliability value of $100\%$ indicates that the algorithm successfully delivers any given message to all sites, ensuring *atomicity* similar to the pure flooding algorithm [26].

**Latency, denoted L** measures the number of hops required to deliver a message to all recipients, i.e., the number of hops of the longest path among all the shortest paths from the source to all other sites that have received the message.

## V. RELIABILITY MODELING

In this section, we model the reliability by analyzing isolated sites in a dissemination graph. We introduce the forwarding probability $p_{forw}$ to uniformly describe retransmission behavior of the different gossip algorithms. According to [23], the message complexity can be easily calculated as function of $p_{forw}$. The latter depends on the parameters of gossip algorithm and the random topologies' characteristic properties, such as high edge dependency or preferential attachment. In Section VI, we access the accuracy of our model and use it to explain our simulation results.

While modeling the reliability of gossip algorithms, we are indifferent to how and when a site is infected. We are only interested, if a site is infected by the end of dissemination. We therefore exploit a *dissemination graph* as follows for modeling the reliability.

Given a graph $G = (S; E)$, and a source site $s_0$, the probabilistic neighbor-aware gossip algorithms follow the same principle: when a new message arrives at a site, this site chooses a subset (possibly empty) of its neighbors in $G$ and forwards the message to these neighbors. We therefore assume, that every site selects a set of neighbors to receive the message in an initial phase. We thus obtain a directed graph $\overrightarrow{G} = (S; \overrightarrow{E})$ with sites identical to $G$. For an edge $(s_i \sim s_j) \in E$, the arc $\overrightarrow{s_is_j}$ is a part of $\overrightarrow{G}$, if $s_i$ has selected $s_j$ in the initial phase. We call $\overrightarrow{G}$ **Dissemination Graph**. Then, a site $s_i$ is infected during dissemination, if there is a path from $s_0$ to $s_i$ in $\overrightarrow{G}$. Furthermore, if every site receives the message, the number of arcs in $\overrightarrow{G}$ expresses the message complexity.

We call a site **isolated** in the dissemination graph, if it has no incoming arcs. Trivially, an isolated site is not infected by a given gossip algorithm. We call the probability that an edge in $E$ becomes an arc towards one site in $S$ the **forwarding probability**, denoted $p_{forw}$.

We assume as a precondition that the random network $G = (S; E)$ is connected. Thus, any message will reach all sites by *pure flooding*. We model the reliability as the probability $1 - P_{iso}$ that no site is isolated in the dissemination graph. Hence, for a random topology with $N$ sites and degree distribution $P(k)$ we model the probability $R$ as

$$R \approx 1 - P_{iso} = \prod_{i=1}^{N-1} \left(1 - (1 - p_{forw}(i))^i\right)^{P(i) \cdot N} \tag{1}$$

where $p_{forw}(i)$ denotes the probability that a site of degree $i$ has an incoming arc from a neighbor. Our evaluation shows that this equation provides an accurate estimate for the reliability and that, due to the degree aware forwarding probability $p_{forw}(i)$, it is applicable to all algorithms and topologies in our study. It is especially suited to predict how reliability differs amongst the studied gossip algorithms and which is more or less reliable at a message complexity that can be obtained by

$$M = \sum_{k=1}^{N-1} P(k) \cdot k \cdot p_{forw}(k) \tag{2}$$

when relatively high reliability is reached [23]. Evidently, the increase of $p_{forw}(k)$ can result in the growth of both reliability and message complexity.

Our model, as presented in Equation (1), assumes that different sites are isolated independently. This holds for *GossipPE*, *GossipPINE*. It does not hold for *GossipFF* and *GossipPISB*, where the events $\overrightarrow{s_i s_j} \in \overrightarrow{E}$ and $\overrightarrow{s_i s_k} \in \overrightarrow{E}$ are not independent. Whether $s_i$ sends to neighbor $s_j$ is influenced, if $s_i$ sends to $s_k$. However, we are mainly interested in relatively high values of message complexity, which provide non-zero reliability. Previous studies [23] have shown that, in these cases only single and widespread sites are isolated. It is thus reasonable to assume independent isolation also for *GossipFF* and *GossipPISB*.

*GossipFF* allows a site to be selected in a neighbor's forwarding list with a probability equal to $fanout$ divided by the neighbor's degree. If $fanout = c_v$, that is exactly what *GossipPISB* does. Thereby, their forwarding probability are identical ($p_{forw}^{FF} = p_{forw}^{PISB} = p_{forw}^{Duo}$) and Equation (1) only gives one model for the two algorithms. Our evaluation shows that on topologies where Equation (1) accurately models the Reliability of *GossipPISB*, this reliability is equal to that of *GossipFF*.

## A. Isolated Sites

In the following, we argue why and when the reliability can be modeled as the probability of an isolated site, as done in Equation (1). In [6], this is shown for large Bernoulli graphs and a dissemination model, encompassing *GossipPE*, *GossipPISB*, and *GossipFF*.

If a site $s_i$ is isolated in the dissemination graph $\overrightarrow{G} = (S; \overrightarrow{E})$, then there exists a cut set $B$ of edges in $E$, separating $s_i$ and the source site $s_0$, such that for any edge $(x \sim y)$ in $B$ where $x$ is a part of the component containing $s_0$, $\overrightarrow{xy}$ is not a part of $\overrightarrow{E}$. If $B$ is minimal, we call $B$ a **border**. The sites adjacent to $B$ which are part of the component containing $s_0$ are called **border-sites**.

Let $P_B$ denote the probability that $\overrightarrow{G}$ contains a border. Clearly $R = 1 - P_B$. Theorem 1 shows that, if a border of size smaller than $k$ forms with non-zero probability, then this probability dominates $P_B$. For *GossipPE*, we can choose $k$ equal to the smallest degree of a site. For *GossipFF*, *GossipPISB*, and *GossipPINE* with high message complexity, borders between sites with low degree cannot form. Then, it can be necessary to choose a larger $k$.

**Theorem 1.** *Let $P_{[B<k]}$ denote the probability that a border $B$ with size $|B| < k$ exists. If $P_{[B<k]} > 0$ for all values of $M < \bar{V}$, then $\lim_{M \to \bar{V}} \frac{P_B}{P_{[B<k]}} = 1$*

If $P_{[iso<k]}$ is the probability, that a site with degree smaller than $k$ is isolated, clearly $R = (1 - P_B) \leqslant (1 - P_{[B<k]}) \leqslant (1 - P_{[iso<k]})$ holds. Based on Theorem 1 we can assume $1 - P_B \approx 1 - P_{[B<k]}$. On a topology with small clustering coefficient, we can further assume that all small borders are isolating single sites, thus $1 - P_{[B<k]} \approx 1 - P_{[iso<k]}$. Since $R \leqslant (1 - P_{iso}) \leqslant (1 - P_{[iso<k]})$ and $R \approx (1 - P_{[iso<k]})$ the approximation in Equation (1) holds. Our model thus gives an exact estimate for the reliability on $\mathcal{B}(N, p_N)$ and $\mathcal{S}(N, m)$ with insignificant clustering coefficient. For $\mathcal{G}(N, \rho)$, the inequalities above imply that our model slightly overestimates the reliability, since some small borders are not considered. This phenomenon will be discussed in Section VI. However, as proven in [37] in a Geometric Graph of high density, the probability that a site does not belong to a *giant component* is dominated by the probability of the site to be isolated. We aspire in our future work to adapt this result to a directional graph, proving that our model is also a good approximation on $\mathcal{G}(N, \rho)$.

### B. Forwarding Probability

In this section, we will deduce the forwarding probability $p_{forw}^{Duo}$ for *GossipFF* and *GossipPISB* depending on the neighbors' degree, whereas we can trivially have $p_{forw}^{PE} = p_e$ and $p_{forw}^{PINE}(k) = \min\{\frac{c_e}{k}, 1\}$ for *GossipPE* and *GossipPINE* respectively on all topologies. Due to the lack of space here, their proofs are shown in Appendix A.

**Lemma 2.** *In $\mathcal{B}(N, p_N)$ if $fanout = c_v = c_{Duo}$, we have $p_{forw}^{Duo}(k) = \frac{\sum_{k<N-1} P(k) \cdot \min\{c_{Duo}, k\}}{\bar{V}}$.*

In $\mathcal{G}(N, \rho)$, before calculating $p_{forw}^{Duo}$ we have to express the dependency of neighbors' degrees. Lemma 3 studies the mean degree of $s_t$, given the degree of its neighbor $s_i$, ignoring the border effect of the rectangular plane.

**Lemma 3.** *In $\mathcal{G}(N, \rho)$, given a site $s_i$ with degree $V_i$, the average degree of a neighbor of $V_i$ is given as: $NeighD = 1 + (V_i - 1) \cdot C + (\bar{V}_s - 1) \cdot (1 - C)$ where $\bar{V}_s = \frac{N \cdot \pi \cdot \rho^2}{a \cdot b}$ is the mean degree of the graph when the border effect of the rectangular region is not taken into account.*

Similar to Lemma 2, with Lemma 3 we can derive $p_{forw}^{Duo}(k) = \min\{\frac{\sum_{k<N-1} P(k) \cdot \min\{c_{Duo}, k\}}{1 + (k-1) \cdot C + (\bar{V}_s - 1) \cdot (1 - C)}, 1\}$, if $fanout = c_e = c_{Duo}$.

In the scale-free graph $\mathcal{S}(N, m)$, generated by the Barabási-Albert model, the edge dependency is very low. Therefore, as argued in Section V-A, it is sound to consider merely the borders that single isolated sites compose and apply Equation (1). However, on account of the *preferential attachment* (see Section II), the degree of two adjacent sites in $\mathcal{S}(N, m)$ is dependent. For example, a site with degree $m$ is never connected to another site with the same degree. Lemma 4 shows the mean degree of neighbors of a site whose degree is $m$ in the graph.

**Lemma 4.** *In $\mathcal{S}(N, m)$, the mean degree of sites that are adjacent to a site with degree $m$ is given by:* $(NeighD + 1) \dfrac{\sum\limits_{s=m_0+1}^{N} \left(\frac{N}{s}\right)^{1/2} \prod\limits_{i>s}^{N} (1-\frac{1}{2i})^m}{\frac{2N}{(m+2)}}$ *where* $NeighD = \sum\limits_{k=m}^{N-1} \dfrac{k(m+1)}{(k+1)(k+2)}$.

We can also study a site with higher degree, for instance $m+1$ in the same way. Nevertheless, this will become more complex, since it should take into account not only the moment when the site joined to the system, but also another moment when another site connects to it. We should point out that the isolation effect brought by sites with degree $m$ dominates the total isolation probability for both *GossipFF* and *GossipPINE*.

Interestingly, we can observe that due to the preferential attachment behavior, a site with low degree has neighbors whose mean degree is higher than the average. For example, given a graph $\mathcal{S}(N, m)$ with $N = 1000$ and $m = 7$, while the average degree amongst an arbitrary site's neighbors is $37$, the neighbors of a site with degree $m(= 7)$ have, in regard to Lemma 4, the mean degree $43$.

Furthermore, *GossipFF* and *GossipPISB* have message retransmission probability that is inversely proportional to the degree of the forwarder. Intuitively, the message reception probability in a site with degree $m$ from an adjacent site is thereby smaller than the constant probability $p_e$ of *GossipPE* with the same message complexity. The probability that a site with degree $m$ receives the message from one neighbor is given by Lemma 5.

**Lemma 5.** *In $\mathcal{S}(N, m)$, for GossipFF and GossipPISB, the probability that one neighbor of a site $s_i$ with degree $m$ forwards the message to $s_i$ is given by:* $p_{forw}^{Duo}(m) = \sum\limits_{k=m+1}^{\lfloor \varphi/Fac \rfloor} \dfrac{m+1}{k(k+1)} + \sum\limits_{k>\varphi/Fac}^{N-1} \dfrac{\varphi(m+1)}{k^2(k+1)Fac}$ *and* $Fac := \dfrac{\sum\limits_{s=m_0}^{N} \left(\frac{N}{s}\right)^{1/2} \prod\limits_{i>s}^{N} (1-\frac{1}{2i})^m}{\frac{2N}{(m+2)}}$ *where for GossipFF,* $\varphi = \min\{fanout, m\}$ *and for GossipPISB,* $\varphi = \min\{c_v, m\}$.

Therefore, for *GossipFF* and *GossipPISB* we restrict Equation (1) to the sites whose degree is $m$ and model the reliability as

$$(1 - (1 - p_{forw})^m)^{p(m) \cdot N} \tag{3}$$

Note that, as we have shown above, sites with degree $m$ have neighbors with high degree. Thereby, the probability for a site whose degree is $m$ to be isolated is larger than $0$ even when a great message complexity is entailed. Hence, Theorem 1 is applicable to *GossipFF* and *GossipPISB* in $\mathcal{S}(N, m)$ with $k = m$. Therefore, Equation (3) gives a good approximate for these algorithms.

The numerical results of our modeling will be addressed along with our simulation in Section VI. We can find our gossip algorithm *GossipPINE* substantially more reliable than the others.

## VI. Performance Evaluation

In this section, we evaluate the metrics that are defined in Section IV. We simulate every probabilistic neighbor-aware gossip algorithm described in Sections III over the random topologies presented in Section II. Our simulations are conducted on top of OMNET++.

We also compare the results, obtained by our reliability modeling, with our simulation. We can observe that in terms of reliability our algorithm outperforms the other gossip algorithms while, in particular, its reliability is even higher in $\mathcal{S}(N,m)$ than in the other random topologies. Furthermore, our reliability modeling presents a good estimate, which establishes the relation between the input parameters for a gossip algorithm and a required reliability.

We consider a network composed of $N = 1000$ sites and, in order to ensure connectivity, $\varepsilon = 1$ for $\mathcal{B}(N, p_N)$ and $\mathcal{G}(N, \rho)$. Since we aim at having almost the same mean degree for all topologies ($\bar{V} \approx 14.0$), the topology parameters were chosen as shown in Table I:

| TOPOLOGY | PARAMETERS |
|---|---|
| $\mathcal{B}(N, p_N)$ | $p_N = 0.014$ |
| $\mathcal{G}(N, \rho)$ | $a = 7500$, $b = 3000$, $\rho = 330$ |
| $\mathcal{S}(N, m)$ | $m_0 = 9$ ($m_0 - clique$), $m = 7$ |

Table I
TOPOLOGY PARAMETERS

For each gossip algorithm, we fixed the parameter values to reach a given message complexity $M$, using the method introduced in [23], and then evaluated the reliability and the latency of the algorithm. 200 different messages are generated by 200 different sources that are chosen uniformly amongst 1000 sites over 50 different graphs for each of the topologies. Then, the results for each message complexity are averaged by the $200 \times 50 = 10000$ message disseminations.

The results are plotted in Figure 1, where the models for *GossipPINE* and *GossipPE* are called **ModelPINE** and **ModelPE** respectively. As explained in Section V, our model produces the same result for *GossipFF* and *GossipPISB*, depicted as **ModelFF**. For a given $p_{forw}$, we can theoretically have both the reliability by our model and the message complexity $M$ shown in Equation (2).

As a matter of fact, the topology parameters can be configured with other values, while similar results are obtained, which verifies our modeling and the best reliability of our algorithm *GossipPINE*. Due to lack of space in our paper, we do not give exhaustive presentation.

**Reliability in $\mathcal{B}(N, p_N)$.** In Figure 1(a), we observe that *GossipPINE* outperforms the other algorithms. Compared to the other algorithms that show equivalent performance, *GossipPINE* achieves the same reliability with roughly 2 fewer messages per site, in accordance with our model.

As a matter of fact, *GossipPINE* increases the message reception in sites with low degree, which are the most susceptible sites to be isolated in the dissemination graph. On the other hand, it decreases redundant message copies in sites with high degree, which reduces message complexity.

**Reliability in $\mathcal{G}(N, \rho)$.** *GossipPINE* also exhibits the best performance in Figure 1(b). Different from $\mathcal{B}(N, p_N)$, the message complexity that is required to achieve high reliability for the other algorithms varies significantly. Thus, *GossipPE* now is far less reliable than the other algorithms. Compared with $\mathcal{B}(N, p_N)$, the boundaries of the rectangular field compose a large number of sites with small degree in $\mathcal{G}(N, \rho)$ (see Section II). These sites are easily isolated in *GossipPE*.

As shown in Lemma 3, in $\mathcal{G}(N, \rho)$ sites with low degree are very likely to be clustered together near the boundaries. This effect is further reinforced by the border of the rectangular region, where many sites with low degree lie. *GossipFF* and *GossipPISB* leverage this clustering, so that the forwarding probability $p_{forw}^{Duo}(i)$ for small $i$ is larger than $p_{forw}^{PE}$. They thus avoid isolating sites in a cluster with low degree and achieve higher reliability than *GossipPE*. Since sites with low degree build clusters, $\mathcal{G}(N, \rho)$ has significantly many small borders that do not isolate a single site, but a group of sites with low degree. The probability for these borders is not included in our model.

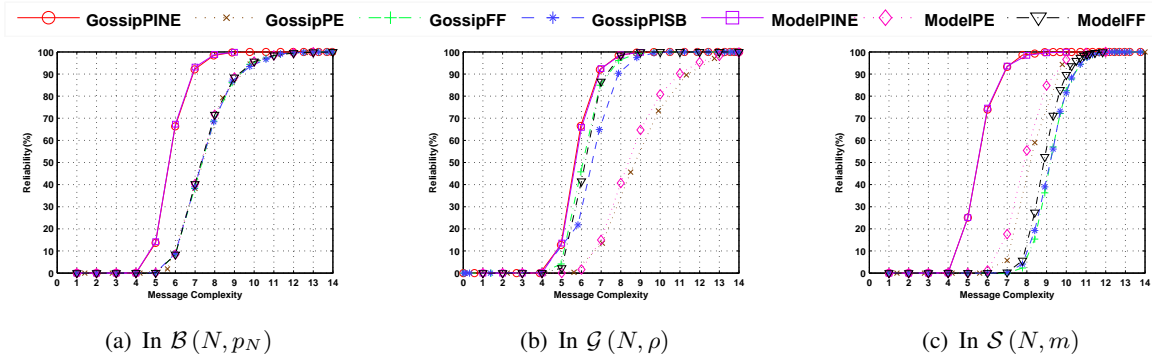

| (a) In $\mathcal{B}(N, p_N)$ | (b) In $\mathcal{G}(N, \rho)$ | (c) In $\mathcal{S}(N, m)$ |

Figure 1. Reliability comparison of gossip algorithms

The results of our modeling and simulation are compared in Figure 1(b). As predicted in Section V-A our model overestimates the reliability for *GossipPE* and *GossipPISB*. However, the predictions for *GossipPE* is still quite accurate. The reliability of *GossipFF* is underestimated, because sites with small degree are often clustered together. Isolation of these sites is not independent, as assumed in Equation (1). The remarkable divergence between *GossipPISB* and *ModelFF* is brought by the high probability for the message to stay within one cluster. The same effect was observed for site percolation in [23]. Note that the probability for a specific border differs in *GossipFF* and *GossipPISB*, if the number of border sites is smaller than the number of edges.

Finally, it is worth pointing out that in *GossipPINE* the probability for a message to be spread out of a cluster with high degree sites and into a cluster with low degree sites is larger than in *GossipPISB* and *GossipPE*. As expected, *GossipPINE* is also accurately described by our model.

**Reliability in** $\mathcal{S}(N, m)$**.** The higher reliability of our algorithm is especially significant for $\mathcal{S}(N, m)$ (see Figure 1(c)). It performs even better in $\mathcal{S}(N, m)$ than on the other topologies.

*GossipFF* and *GossipPISB* perform worst, on account of their poor exploitation of hubs. More precisely, a site with low degree is more probably connected to a hub due to the *preferential attachment*. As proven by Lemma 4, the neighbors of sites with low degree have a degree above average. Since hubs have a lower forwarding probability in *GossipFF* and *GossipPISB* the probability to isolate sites with low degree is higher than in the other algorithms. In our experimental setup, approximately 222 sites have the lowest degree 7. Calculating the isolation probability for only these sites, as done for **ModelFF**, using Equation (3), already gives a reasonable estimate. The reliability for *GossipFF* and *GossipPISB* is identical, verifying the correctness of their interchangeable reliability modeling. *GossipPINE* and *GossipPE* are precisely modeled by Equation (1).

**Latency:** An efficient gossip algorithm should not only maximize reliability but also mini-mize latency. We did record the latency during our simulations and verified that in $\mathcal{B}(N, p_N)$ and $\mathcal{G}(N, \rho)$ latency of the different algorithms for the same message complexity differs only little, whereas *GossipPINE* shows the smaller latency. In $\mathcal{S}(N, m)$ there exist some values for message complexity, where *GossipPINE* has a significantly smaller latency than *GossipFF* and *GossipPISB*. However, for just single values, *GossipPE* shows even a smaller latency than *GossipPINE*, while this difference of $0.5\%$ is insignificant. Results are addressed in Appendix A.

## VII. RELATED WORK

Reliable information dissemination is essential for distributed systems and applications, where it is hard to obtain a global view of the network. Thereby, many existing works aim at proposing gossip algorithms with high reliability at as little message complexity as possible, exploiting a one-hop neighborhood information in every site. A reliability model is thus necessary to compare different algorithms and show how to fix input parameters for the gossip algorithms. An evaluation similar to the ours was done in [23], but without regarding *GossipPINE* and offering no theoretical analysis.

A reliability model for *GossipFF* is established in [26] over peer-to-peer systems by using random graph theory [10], while this model explains a $fanout$ adaptive protocol in [15] for a heterogeneous network. The reliability is also analyzed by Markov Chain model in [12] over a complete graph, whereas a stochastic model is tailored in [31] to adapt ad-hoc network. However, the models applied in this work do not extend to more general situations, such as our algorithm *GossipPINE* that is more efficient than *GossipFF* on every random topology in our study.

In [20], bond percolation and site percolation are studied in random graphs. The former matches *GossipPE*, which never reaches a worse reliability than a probabilistic broadcast algorithm in [21] that is modeled by the latter, as shown in [40]. Nevertheless, unlike our reliability model, it cannot explain how to fine-tune the parameter in a gossip algorithm to reach a desirable reliability, nor does the result extend to other gossip algorithms.

A modified version of *GossipPE* is analyzed over $\mathcal{B}(N, p_N)$ and $\mathcal{S}(N, m)$ by a model proposed in [33], which generalizes two common models: The Maki-Thompson model [32] and the Susceptible-Infected-Removed model. The impact of degree distribution on information spreading is addressed, while they did not discuss the edge dependency, which cannot be neglected for instance in $\mathcal{G}(N, \rho)$.

In [41], the authors proposed a neighbor degree-based algorithm, where the transmission probability towards each neighbor of a site is reversely related to the minimum value of its degree and that of its neighbors. Their algorithm is thus a combination of the algorithms *GossipPISB* and *GossipPINE* studied by us. They theoretically evaluate its reliability, which is confirmed by simulations in $\mathcal{B}(N, p_N)$ and a scale-free network without preferential attachment. Nonetheless, we showed that *GossipPISB* gives a poor reliability in a graph with preferential attachment and is difficult to model in graphs with dependency on neighbor's degree. Furthermore, our algorithm *GossipPINE* stands out by its simplicity, that makes it both easy to model and deploy. In [7], the authors also specify forwarding probability of a site in ad-hoc networks as function of both its degree and the mean degree of its neighbors, while the reliability is heuristically studied. The Smart Gossip protocol in [28] being aware of edge dependency constructs local relationship trees in two-hop neighborhoods, which allows to decide an optimum forwarding probability.

Maintenance of the relationship tree is required. In comparison, our algorithm is efficient and very simple to apply in every network.

The authors distinguish all sites by four levels in [2]. Then, the probability in each level is reversely adapted to the number of levels predefined in sensor network. This ensures that sites in a sparse density area retransmit the message with a higher probability. However, as we analyzed, for $\mathcal{S}(N, m)$ such an idea is not suitable at all, since the hubs' dissemination power is restricted.

In order to obtain $100\%$ of reliability, some protocols add a pull phase in [16] or exploit specific random graph properties of $\mathcal{S}(N, m)$ [24]. Even so, the pull phase is difficult to implement [26], while hubs and the preferential attachment do not exist in the random graphs $\mathcal{B}(N, p_N)$ and $\mathcal{G}(N, \rho)$.

## VIII. CONCLUSION

In this article, we have proposed a new algorithm and a model for the reliability of neighbor-aware gossip algorithms in random networks. Our model accurately predicts the reliability of different algorithms over various random topologies as shown by our simulations.

Our algorithm *GossipPINE* outperforms three common gossip algorithms in every random graph in terms of reliability. This is confirmed both by our model and simulations. Furthermore, we have shown that the reliability of *GossipPINE* does not depend on more complex properties of the topology, such as the dependency between the degrees of adjacent sites. It is therefore applicable even in networks, whose global characteristics can only be estimated.

## APPENDIX

**Theorem 1.** *Let $P_{[B<k]}$ denote the probability that a border $B$ with size $|B| < k$ exists. If $P_{[B<k]} > 0$ for all values of $M < \bar{V}$, then*

$$\lim_{M \to \bar{V}} \frac{P_B}{P_{[B<k]}} = 1.$$

*Proof.* For *GossipPE*, we consider $M \to \bar{V}$. Thereby, $p_e \to 1$ and we denote $q$ for $1 - p_e$. Let $B \cap [B < k]^c$ describe the event that there is a border but no border with size smaller than $k$. Since $P_B = P_{B \cap [B<k]^c} + P_{[B<k]}$, it is enough to show that $\frac{P_{B \cap [B<k]^c}}{P_{[B<k]}}$ tends to zero. Since $P_{[B<k]} > 0$, there exists a possible border $B_i$ with $|B_i| < k$, and $\frac{P_{B \cap [B<k]^c}}{P_{[B<k]}} \leqslant \frac{P_{[B \geqslant k]}}{P_{[B<k]}} \leqslant \frac{P_{[B \geqslant k]}}{P_{B_i}} \leqslant \frac{P_{[B \geqslant k]}}{q^{k-1}}$. We denote $P_{[B \geqslant k]}$ the sum over all possible dissemination graphs with a border whose size is at least $k$, which are denoted $\overrightarrow{G_B}$. Therefore, $P_{[B \geqslant k]} = \sum\limits_{\overrightarrow{G_B}} p_e^i q^j$. Since all graphs $\overrightarrow{G_B}$ have at least $k$ arcs that are not used, and $p_e = 1 - q$, we can find a polynomial $Q$, such that $P_{[B \geqslant k]} = q^k Q(q)$. Thus, $\lim_{q \to 0} \frac{q^k Q(q)}{q^{k-1}} = 0$.

When choosing the border $B_i$ with size smaller than $k$, we have to make sure that $P_{B_i} > 0$ for all $M < \bar{V}$. The theorem follows for the rest of the gossip algorithms, since for any border $B_j$ with size $k$ or larger, $P_{B_j}$ decreases faster than $P_{B_i}$.

Note that $k$ will be quite large for *GossipPINE*, since the degree of a site isolated by a border of size $k$ has to be larger than $c_e$. □

**Lemma 2.** *In $\mathcal{B}(N, p_N)$ if $fanout = c_v = c_{Duo}$, we have*

$$p_{forw}^{Duo}(k) = \frac{\sum\limits_{k < N-1} P(k) \cdot \min\{c_{Duo}, k\}}{\bar{V}}.$$

*Proof.* For *GossipFF* and *GossipPISB*, the forwarding probability to receive an arc from its neighbor depends on the neighbor's degree. Since $\mathcal{B}(N, p_N)$ is an uncorrelated network [33], the neighbors' degree is independent of $k$. Thereby, the forwarding probability is simply the probability for one arc to appear in $\overrightarrow{G}$. For a site, according to [23] if $fanout = c_v = c_{Duo}$, then $\sum\limits_{k=1}^{N-1} P(k) \cdot \min\{c_{Duo}, k\}$ gives the message complexity or the average number of outgoing arcs resulted from both gossip algorithms, while $\bar{V}$ gives the average number of possible outgoing arcs of the dissemination graph. Therefore, $p_{forw}^{Duo}$ is the ratio between them. □

**Lemma 3.** *In $\mathcal{G}(N, \rho)$, given a site $s_i$ with degree $V_i$, the average degree of a neighbor of $V_i$ is given as:*

$$NeighD = 1 + (V_i - 1) \cdot C + (\bar{V}_s - 1) \cdot (1 - C)$$

*where $\bar{V}_s = \frac{N \cdot \pi \cdot \rho^2}{a \cdot b}$ is the mean degree of the graph when the border effect of the rectangular region is not taken into account.*

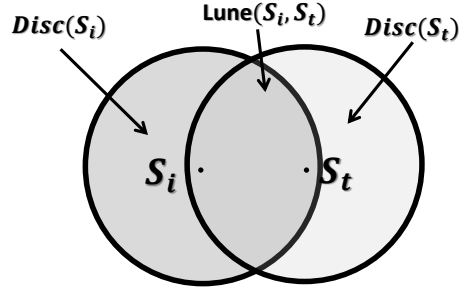

Figure 2. Two kinds of neighbors

*Proof.* Figure 2 shows $s_i$, a neighbor $s_t$, and the retransmission zones (i.e., $Disc(s_t)$, $Disc(s_i)$). We call the overlapped part between two discs **lune** (i.e., $Lune(s_i, s_t)$). Let *disc* be the area of $Disc(s_t)$. Then, the average area of the lune is $C \cdot disc$. Clearly, the density of sites inside $Disc(s_i)$, and also inside $Lune(s_i, s_t)$ is $\frac{V_i}{disc}$. If *disc* is small, we can assume that the density outside $Disc(s_i)$ is $\frac{\bar{V}}{disc}$. Therefore, $s_t$ has $\frac{V_i}{disc} \cdot C \cdot disc$ neighbors in $Lune(s_i, s_t)$, $\frac{\bar{V}}{disc} \cdot (1-C) \cdot disc$ neighbors outside $Lune(s_i, s_t)$, plus $s_i$. □

**Lemma 4.** *In $\mathcal{S}(N, m)$, the mean degree of sites that are adjacent to a site with degree $m$ is given by:*

$$(NeighD + 1)\frac{\sum\limits_{s=m_0+1}^{N} \left(\frac{N}{s}\right)^{1/2} \prod\limits_{i>s}^{N}(1 - \frac{1}{2i})^m}{\frac{2N}{(m+2)}}$$

*where* $NeighD = \sum\limits_{k=m}^{N-1} \frac{k(m+1)}{(k+1)(k+2)}$.

*Proof.* For an arbitrary site $s_i$ adjacent to site $s_t$ in $\mathcal{S}(N, m)$, as shown in [24], the degree distribution for $s_t$ with degree $k$ is $\frac{k \cdot P(k)}{2 \cdot m}$. Thus, the average degree amongst neighbors of $s_i$ is $NeighD = \sum\limits_{k=m}^{N-1} \frac{k^2 \cdot P(k)}{2 \cdot m} = \sum\limits_{k=m}^{N-1} \frac{k(m+1)}{(k+1)(k+2)}$.

Now consider a site $s_i$ whose degree is $V_i = m$. According to the generation of $\mathcal{S}(N, m)$ shown in Section II, the neighbors of $s_i$ are only those chosen when $s_i$ is added to the network.

Assume that $s_i$ is the last site that comes to the graph. While $s_i$ connects to a site whose degree is $k$ with probability $\frac{k \cdot p(k)}{2 \cdot m}$, the average degree of sites adjacent to $s_i$ is in turn increased by one (i.e., $NeighD + 1$).

If we do not assume that $s_i$ is the last added site, the mean degree of its neighbors is even higher, since their degree might have increased by adding more sites to the system. It is shown in [35] that when $q$ sites are added to a graph $\mathcal{S}(N, m)$ with $n$ sites, the average increase factor of degree for any of the initial sites is given by $\left(\frac{q+n}{n}\right)^{1/2}$. In the following analysis, since the total number of sites $N$ is much larger than the size of initial cliques ($N \gg m_0$), we ignore the initial $m_0$ sites. Given a natural number $s$ smaller than $N$, we call the $s$th site that is added during the graph generation a site with age $N - s$.

Trivially, the probability that a site $s_i$ with degree $m$ has age $N - s$ is the probability that no younger site is connected to $s_i$. This is given by: $\prod_{i>s}^{N}(1 - \frac{1}{2i})^m$. Thereby, the probability for a site to have both degree $m$ and age $N - s$ is $\frac{\prod_{i>s}^{N}(1-\frac{1}{2i})^m}{2N/(m+2)}$. Therefore, given a site with degree $m$, the mean degree of its neighbors is:

$$(NeighD + 1)\frac{\sum_{s=m_0+1}^{N}\left(\frac{N}{s}\right)^{1/2}\prod_{i>s}^{N}(1-\frac{1}{2i})^m}{\frac{2N}{(m+2)}}.$$

□

**Lemma 5.** *In $\mathcal{S}(N, m)$, for GossipFF and GossipPISB, the probability that one neighbor of a site $s_i$ with degree $m$ forwards the message to $s_i$ is given by:*

$$p_{forw}^{Duo}(m) = \sum_{k=m+1}^{\lfloor \varphi/Fac \rfloor} \frac{m+1}{k(k+1)} + \sum_{k>\varphi/Fac}^{N-1} \frac{\varphi(m+1)}{k^2(k+1)Fac}$$

*and*

$$Fac := \frac{\sum_{s=m_0}^{N}\left(\frac{N}{s}\right)^{1/2}\prod_{i>s}^{N}(1-\frac{1}{2i})^m}{\frac{2N}{(m+2)}}$$

*where for GossipFF, $\varphi = \min\{fanout, m\}$ and for GossipPISB, $\varphi = \min\{c_v, m\}$.*

*Proof.* Let $s_i$ be a site of degree $m$. Following Lemma 4, we can find the degree distribution amongst $s_i$'s neighbors, which is their initial degree just before $s_i$ joins the network, multiplied by a factor which is as function of the age of $s_i$.

By analogy, the average of this age-factor is given by

$$Fac := \frac{\sum_{s=1}^{N}\left(\frac{N}{s}\right)^{1/2}\prod_{i>s}^{N}(1-\frac{1}{2i})^m}{\frac{2N}{(m+2)}}$$

Assuming that the initial degree distribution among neighbors of $s_i$ is independent of $s_i$'s age, we thus have the degree of every neighbor of $s$ multiplied by factor $Fac$. Moreover, notice that the probability for $s_i$ to be connected to a site with initial degree $k$ is **0** if $k \leqslant m$, and otherwise:

$$P_{N_s-init}(k) := \frac{(k-1) \cdot P(k-1)}{2m}$$

$$= \frac{m+1}{k(k+1)}$$

Then, given $\varphi = \min\{fanout, m\}$ for *GossipFF* and $\varphi = \min\{c_v, m\}$ for *GossipPISB*, the probability that a neighbor forwards the message to $s_i$ can be approximated by

$$p_{forw}^{Duo} = \sum_{k \cdot Fac \leqslant \varphi} P_{N_s-init}(k) + \sum_{k \cdot Fac > \varphi}^{N-1} \frac{\varphi}{k \cdot Fac} P_{N_s-init}(k)$$

where the first term represents the probability to receive the message from an adjacent site with degree lower than $\varphi$, while the second gives the complementary case. □

An efficient gossip algorithm should not only maximize reliability but also minimize latency with low message complexity. We analyze the latency performance when the reliability reaches at least $85\%$. Figures 3(a), 3(b), and 3(c) show the simulation results in $\mathcal{B}(N, p_N)$, $\mathcal{G}(N, \rho)$, and $\mathcal{S}(N, m)$ respectively.

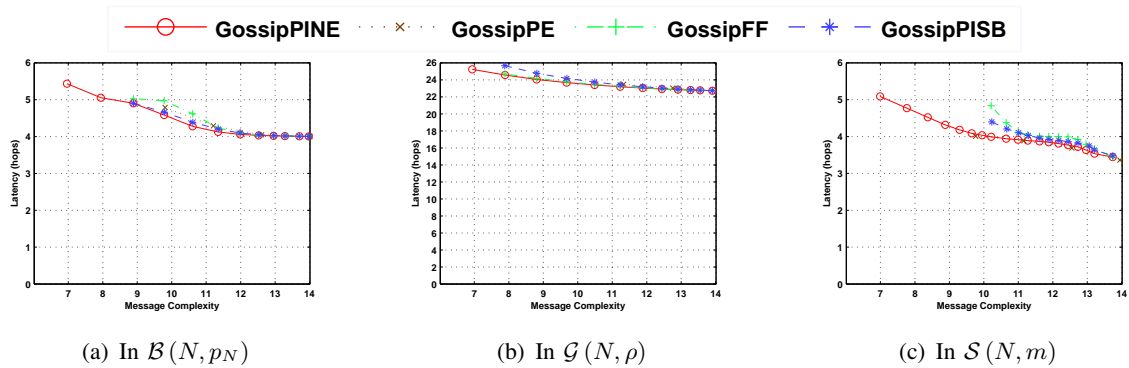

(a) In $\mathcal{B}(N, p_N)$  (b) In $\mathcal{G}(N, \rho)$  (c) In $\mathcal{S}(N, m)$

Figure 3. Latency comparison of gossip algorithms

Over all random topologies, after a given message complexity, latency does no longer decrease, but converges towards the pure flooding approach (i.e., the shortest routes between the source and the other sites), and therefore, towards the minimum latency.

In $\mathcal{B}(N, p_N)$, the latency of our algorithm *GossipPINE* reaches the minimum value (i.e., 4 hops) with the smallest message complexity amongst all gossip algorithms. However, the difference is very small. A similar behavior of the latency curves is observed in $\mathcal{G}(N, \rho)$, except that the minimum latency value lies around 22 hops, since the diameter of $\mathcal{G}(N, \rho)$ is greater than that of $\mathcal{B}(N, p_N)$.

For $\mathcal{S}(N, m)$, although the latency of all gossip algorithms converges to a smaller value than for the previous networks, the performance difference becomes significant. It can be explained by the exploitation of the dissemination potential of hubs. As a matter of fact, both *GossipFF* and *GossipPISB* limit the dissemination power (i.e., the number of neighbors to receive the message under the pure flooding) of sites with high degree, and thus discard numerous short-cut paths. On the contrary, *GossipPE* best distributes such power to the sites whose degree is high, such as hubs that are connected with high probability, which constitute the heart of the network [24]. *GossipPINE* sometimes fails to use the shortest path towards hubs. Thus, for example when $M = 10$, our algorithm slightly under-performs *GossipPE*, which is $0.5\%$ less effective. Even so, *GossipPINE* gains over $5\%$ higher reliability than *GossipPE* at $M = 10$. Furthermore, for $M > 10.5$, they show the same latency, while *GossipPINE* always proposes a better reliability.

## REFERENCES

[1] Omnet++: Discrete event simulation system *http://www.omnetpp.org/*.
[2] J.-D. Abdulai, A. Mohammed, K. S. Nokoe, and E. Oyetunji. Route discovery in wireless mobile ad hoc networks with adjusted probabilistic flooding. adaptive science technology. *ICAST*, 2009.:99–109, 2009.
[3] R. Albert and A.-L. Barabási. Statistical mechanics of complex networks. *Reviews of Modern Physics*, 74:47–97, Jan. 2002.
[4] X. An and R. Hekmat. Probabilistic-based message dissemination in ad-hoc and sensor networks using directional antennas. *International Conference on Mobile Adhoc and Sensor Systems*, pages 432–438, 2009.
[5] C. Avin. Distance graphs: From random geometric graphs to bernoulli graphs and between. *DIALM-POMC*, pages 71–78, 2008.
[6] F. Ball and A. D. Barbour. Poisson approximation for some epodemic models. *Journal of Applied Probability*, 21(3):479–490, 1990.
[7] J. Cartigny and D. Simplot. Border node retransmission based probalistic broadcast protocols in ad-hoc networks. *Annual Hawaii International Conference on System Sciences*, pages 6–9, Jan. 2003.
[8] B. Doerr, M. Fouz, and T. Friedrich. Why rumors spread so quickly in social networks. *Commun. ACM*, 55(6):70–75, 2012.
[9] V. Drabkin, R. Friedman, G. Kliot, and M. Segal. Rapid: Reliable probabilistic dissemination in wireless ad-hoc networks. Technical report, Computer Science Department, Technion - Israel Institute of Technology, 2006.
[10] P. Erdős and A. Rényi. On random graphs I. *Publicationes Mathematicae*, 6:290–297, 1959.
[11] P. Erdős and A. Rényi. On the evolution of random graphs. *Publ. Math. Inst. Hung. Acad. Sci.*, 5(17):17–60, 1960.
[12] P. T. Eugster, R. Guerraoui, S. B. Handurukande, A.-M. Kermarrec, and P. Kouznetsov. Lightweight probabilistic broadcast. *ACM Transaction on Computer Systems*, 21:341–374, 2003.
[13] P. T. Eugster, R. Guerraoui, and P. Kouznetsov. $\delta$-reliability: A probabilistic measure of broadcast reliability. *International Conference on Distributed Computing Systems*, pages 24–26, Mar. 2004.
[14] P. T. Eugster, R. Guerraoui, A. m. Kermarrec, and L. Massoulié. From epidemics to distributed computing. *IEEE Computer*, 37(5):60–67, May 2004.
[15] D. Frey, R. Guerraouia, A.-M. Kermarrec, B. Koldehofe, M. Mogensen, M. Monod, and V. Quéma. Heterogeneous gossip. *International Conference on Middleware*, pages 42–61, 2009.
[16] R. Friedman, V. Drabkin, G. Kliot, and M. Segal. On reliable dissemination in wireless ad-hoc networks. *IEEE Transactions on Dependable and Secure Computing*, 8(6):866–882, 2011.
[17] R. Friedman and A. C. Viana. Gossiping on manets: The beauty and the beast. *Operating Systems Review*, 41(5):67–74, Oct. 2007.
[18] A. Fronczak, P. Fronczak, and J. A. Hołyst. Mean-field theory for clustering coefficients in barabási-albert networks. *Phys. Rev. E*, 68(4):046126, oct 2003.
[19] B. Garbinato, D. Rochat, and M. Tomassini. Impact of scale-free topologies on gossiping in ad hoc networks. *NCA*, pages 269–272, 2007.
[20] G. Grimmett. *Percolation*. Springer, 1989.
[21] Z. Haas, J. Halpern, and L. Li. Gossip-based ad hoc routing. *INFOCOM*, 3:1707–1716, 2002.
[22] R. Hu. *Epidemic dissemination algorithms in large-scale networks: comparison and adaption to topologies*. PhD thesis, Université de Pierre Marie Curie, Dec. 2013.
[23] R. Hu, J. Sopena, L. Arantes, P. Sens, and I. Demeure. Fair comparison of gossip algorithms over large-scale random topologies. *International Symposium on Reliable Distributed Systems*, pages 331–340, Oct. 2012.
[24] R. Hu, J. Sopena, L. Arantes, P. Sens, and I. Demeure. Efficient dissemination algorithm for scale-free topologies. *International Conference on Parallel Processing*, Oct. 2013.
[25] J. Jetcheva, Y. Hu, D. Maltz, and D. Johnson. A simple protocol for multicast and broadcast in mobile ad hoc networks. *Internet Draft: draft-ietf-manet-simple-mbcast-01.txt*, 2001.
[26] A.-M. Kermarrec, L. Massoulié, and A. Ganesh. Probabilistic reliable dissemination in large-scale systems. *IEEE TPDS*, 3:248–258, Mar. 2003.
[27] Z. Kong and E. M. Yeh. Characterization of the critical density for percolation in random geometric graphs. *ISIT*, 2007.
[28] P. Kyasanur, R. R. Choudhury, and I. Gupta. Smart gossip: An adaptive gossip-based broadcasting service for sensor networks. *International Conference on Mobile Adhoc and Sensor Systems*, pages 91–100, Oct. 2006.
[29] J. Leitao, J. Pereira, and L. Rodrigues. Epidemic broadcast trees. *SRDS*, pages 301–310, 2007.

[30] J. Lipman, P. Boustead, and J. Judge. Neighbor aware adaptive power flooding in mobile ad hoc networks. *International Journal of Foundations of Computer Science*, 14(2):237–252, 2003.

[31] J. Luo, P. T. Eugster, and J.-P. Hubaux. Route driven gossip: Probabilistic reliable multicast in ad-hoc networks. *INFOCOM*, 3:2229–2239, 2003.

[32] D. P. Maki and M. Thompson. Mathematical models and applications: with emphasis on the social, life and management sciences, prentice-hall, englewood cliffs. *N. J.*, 1973.

[33] M. Nekovee, Y. Moreno, G. Bianconi, and M. Marsili. Theory of rumour spreading in complex social networks. *PHYSICA A*, 374:457, 2007.

[34] S.-Y. Ni, Y.-C. Tseng, Y.-S.Chen, and J.-P. Sheu. The broadcast storm problem in a mobile ad hoc network. *International Conference on Mobile Computing and Networking*, 8:151–162, 1999.

[35] R. Pastor-Satorras and A. Vespignani. *Evolution and Structure of Internet: A Statistical Physics Approach*. Cambridge University Press, 2004.

[36] M. Penrose. *Random Geometric Graphs*, volume 5. Oxford University Press, oxford studies in probability edition, May 2003.

[37] M. D. Penrose. On a continuum percolation model. *Advances in Applied Probability*, 23(3):536–556, 1991.

[38] T. Philips, S. Panwar, and A. Tantawi. Connectivity properties of a packet radio network model. *IEEE Transactions on Information Theory*, pages 1044–1047, 1989.

[39] A. Polynikisa. Random walks and scale-free networks. Master's thesis, University of York, UK, 2006.

[40] C.-C. Shen, Z. Huang, and C. Jaikaeo. Directional broadcast for mobile ad hoc networks with percolation theory. *IEEE Transactions on Mobile Computing*, 5(4):317–332, Apr. 2006.

[41] A. O. Stauffer and V. C. Barbosa. Probabilistic heuristics for disseminating information in networks. *IEEE/ACM Transactions on Networking*, 15:425–435, 2007.

[42] D. Ustebay, R. Castro, and M. Rabbat. Efficient decentralized approximation via selective gossip. *Journal of Selected Topics in Signal Processing*, PP(99):1, 2011.